\documentclass[11pt, oneside]{article}

\usepackage[utf8]{inputenc}
\usepackage{amsmath}
\usepackage{graphicx}
\usepackage{hyperref}
\usepackage{booktabs} 
\usepackage{multirow}
\usepackage{array}    
\usepackage[table,xcdraw]{xcolor} 

\usepackage[
    top=2cm,
    bottom=2cm,
    left=2cm,
    right=2cm,
    headheight=17pt, 
    includehead,includefoot,
    heightrounded, 
]{geometry}

\usepackage[backend=biber, style=chicago-authordate, natbib=true]{biblatex}
\begin{document}

\begin{titlepage}
    \centering
    Stevens Institute of Technology \\
    \vspace{1.5cm}
    \vspace{4cm}
    {\huge\bfseries High-Frequency Trading Liquidity Analysis \\}
    \vspace{0.5cm}
    {\Large Application of Machine Learning Classification \\}
    \vspace{0.5cm}
    \vfill
    \textsc{\Large Sid Bhatia, Sidharth Peri, \\ Sam Friedman, Michelle Malen \\}
    \vfill
    {\large August 5, 2024}
\end{titlepage}

\newgeometry{
    margin=0.8in
}

\section{Introduction}

Liquidity is fundamentally intertwined with the dynamics of financial markets and plays a critical role in influencing transaction costs and market stability. A deficiency in liquidity during trading hours can dramatically increase transaction costs, adversely impacting market stability. Utilizing established models, we have derived several liquidity measures from various market characteristics, including tightness, depth, resiliency, and trading dynamics.

In previous research, we focused on liquidity measures derived from trades and the limit order book, effectively clustering relevant liquidity metrics and identifying significant outlier events. This current research builds on these foundations by proposing to develop a robust framework for analyzing liquidity using High-Frequency Trading (HFT) data. The primary goal of this study is to uncover critical insights across multiple domains, including the identification and management of liquidity risk, the creation of statistical models based on liquidity analysis, and the generation of novel inputs for comprehensive financial network evaluations.

\section{Literature Review}

\subsection{Macroeconomic Events and Liquidity}

The impact of macroeconomic events on liquidity has been extensively studied, with significant contributions from researchers such as Kong. Their analysis of liquidity measures during Brexit highlights how geopolitical events can lead to substantial fluctuations in market liquidity, affecting both market stability and efficiency. This study provides a foundation for understanding the broader implications of macroeconomic changes on financial systems.

\subsection{Detection of Rare Events}

The methodology for detecting rare events in financial markets has been advanced by Golbayani and Bozdog, who employed Zonoid depth functions to identify outlier events in financial datasets. Their work is crucial for understanding how extreme market conditions can disrupt market equilibrium and affect liquidity, thereby helping market analysts and traders anticipate and mitigate potential risks.

\subsection{Political Turmoil and Market Liquidity}

Exploring the direct impact of political events on market dynamics, Mago and others investigate liquidity risks and asset movements during the Brexit referendum. Their findings underscore the challenges markets face during periods of political uncertainty and the heightened liquidity risks that can arise, stressing the need for effective risk management strategies to combat these effects.

\subsection{Cluster Analysis in Liquidity Measurement}

Salighehdar and others contribute to liquidity modeling by employing cluster analysis on high-frequency data to study liquidity measures in stock markets. Their research demonstrates that liquidity can be effectively predicted through advanced statistical techniques, providing crucial insights for the development of more resilient financial strategies.

\subsection{Multidimensional Analysis of Liquidity}

Zaika extends the analysis of liquidity measures by focusing on rare events and their multidimensional characteristics. This approach enriches our understanding of liquidity beyond traditional models, offering comprehensive strategies to handle liquidity under diverse market conditions, especially in high-frequency trading scenarios.

\subsection{Geopolitical Impact on Liquidity Distribution}

Further emphasizing the sensitivity of markets to geopolitical changes, Kong and others examine how Brexit influenced liquidity distribution characteristics. This study aligns with other research by illustrating the profound impact external shocks can have on liquidity and the ongoing need for financial markets to adapt to these changes dynamically.

These sections collectively underscore the complexity of liquidity in financial markets, revealing how various factors, from geopolitical to macroeconomic, influence liquidity and necessitate robust analytical frameworks to manage it effectively.

\section{Research Agenda}

\subsection{Study Objectives}

This research agenda involves using limit order book datasets to develop predictive models for price movements, utilizing high-frequency data from the Refinitiv Tick History Dataset. 

\subsection{Planned Activities}

\begin{itemize}
    \item Conduct a comprehensive literature review.
    \item Process extensive datasets, specifically Trade and Quote (TAQ) and Limit Order Book (LOB).
    \item Calculate and analyze various liquidity measures.
    \item Develop predictive models incorporating statistical and machine learning techniques.
\end{itemize}

\section{Methodology \& Data}

\subsection{Data Collection}

The core aim of this project is to utilize TAQ data to compute liquidity measures that serve as predictors in a model designed to forecast price movements. Our primary focus is on the price movements calculated from the trade prices recorded in the TAQ data.

\subsection{Data Sampling and Reduction}

Data sampling includes daily subsets from 11:00 AM to 4:00 PM. This subset is further refined to 1-minute frequency data, providing approximately 300 data points per ticker per day. For each minute, the dataset comprises the first trade and all quotes within that minute. Specific liquidity metrics derived from these minute intervals will form the independent variables of our predictive model.

\section{Predictive Model Development}

\subsection{Model Overview}

The predictive model is approached as a classification problem where the main dependent variable is the price movement direction, categorized as Up or Down. The model will be implemented using the following classification techniques:

\begin{itemize}
    \item Logistic Regression (LR)
    \item Support Vector Machine (SVM)
    \item Random Forest Classifier (RF)
\end{itemize}

\subsection{Model Optimization}

Initially, all models are fitted with all available independent variables. Subsequent subset selection aims to identify the optimal combination of liquidity measures that minimize classification errors. Data will be split in a 70\%-15\%-15\% manner across training, validation, and testing sets respectively.

\section{Model Background and Implementation}

\subsection{Data Filtration and Processing}

Data filtration initially restricts the dataset to entries between 11:00 AM and 4:00 PM, focusing on 1-minute intervals. For each interval, trade data and all quotes for that minute are collected. The average quoted bid and ask prices, as well as sizes, are calculated for each minute.

\subsection{Liquidity Metrics}

A series of liquidity metrics form the independent variables for our model:

\begin{itemize}
    \item Turnover, Market Depth, Log Depth, Dollar Depth
    \item Various spreads including Spread, Effective Spread, and various Relative Spreads
    \item Quote Slope, Log Quote Slope, Adjusted Log Quote Slope
    \item Composite Liquidity, Liquidity Ratio 1 (Amivest), Flow Ratio, Order Ratio, Illiquidity (Amihud)
\end{itemize}

\section{Results}

The following section presents the confusion matrices and accuracy metrics for our three models using baseline features, subset-selected features, and those selected based on feature importance. In the confusion matrices, the top-left cell represents correctly predicted "Up" market movements, while the bottom-right cell represents correctly predicted "Down" movements. The top-right and bottom-left cells indicate misclassifications within the test set. Accuracy is expressed as the percentage of correct predictions regarding market direction. Additionally, we provide feature importance coefficients for the RF and SVM models, along with visual representations of these feature importances.

\vspace{0.5cm}

\begin{table}[h!]
\centering
\begin{tabular}{|l|c|c|c|c|}
\hline
\rowcolor[HTML]{000000} 
\textcolor{white}{\textbf{Model}} & \textcolor{white}{\textbf{All Features}} & \textcolor{white}{\textbf{Accuracy\_AF}} & \textcolor{white}{\textbf{Feature Combination}} & \textcolor{white}{\textbf{Accuracy\_FC}} \\ \hline
\rowcolor[HTML]{C0C0C0} 
LOG & $\begin{bmatrix} 25 & 4 \\ 4 & 7 \end{bmatrix}$ & \textbf{62.75\%} & $\begin{bmatrix} 28 & 1 \\ 22 & 0 \end{bmatrix}$ & 54.90\% \\ \hline
SVM & $\begin{bmatrix} 28 & 1 \\ 18 & 4 \end{bmatrix}$ & \textbf{62.79\%} & $\begin{bmatrix} 16 & 13 \\ 12 & 10 \end{bmatrix}$ & 50.98\% \\ \hline
\rowcolor[HTML]{C0C0C0} 
RF & $\begin{bmatrix} 22 & 7 \\ 13 & 9 \end{bmatrix}$ & 60.78\% & $\begin{bmatrix} 27 & 2 \\ 17 & 5 \end{bmatrix}$ & \textbf{62.74\%} \\ \hline
\end{tabular}
\end{table}

\vspace{0.5cm}

Key insights from these results include the notably high accuracy of the baseline models compared to the tuned models that utilized a subset of liquidity metrics. This outcome suggests that all liquidity metrics may be relevant, and that predicting market direction becomes challenging when using a limited set of these metrics. The analysis of feature importance yields interesting observations: Liquidity Ratio, Flow Ratio, and Turnover consistently emerge as significant across all models. These metrics exhibit the highest feature importance scores, indicating their substantial influence on the models' predictions.

\section{Conclusion}

In this study, we have developed a comprehensive framework for liquidity analysis, underscoring the pivotal role of liquidity in financial markets and its profound impact on transaction costs and market stability. Our research has contributed to the understanding of identifying and regulating liquidity risk, alongside the formulation of statistical models rooted in liquidity analysis.

Through the application of various methodologies, including advanced machine learning algorithms such as Logistic Regression, Support Vector Machine, and Random Forest, we successfully identified critical liquidity metrics. In our efforts to predict return direction at minute intervals, our findings suggest that utilizing an extensive set of liquidity metrics yields more accurate results compared to models with reduced feature sets. This may be attributed to the inherent complexity of making predictions at very short time intervals, which likely necessitates a broader range of inputs for enhanced predictive accuracy.

When examining individual feature importance within our SVM and RF models, Liquidity Ratio, Flow Ratio, and Turnover consistently emerged as the most influential metrics. Notably, the Random Forest algorithm outperformed others, delivering the highest accuracy in our predictions. These findings highlight the significance of these liquidity measures, indicating that variations in these metrics can substantially affect predictive outcomes, further emphasizing their role as key determinants of liquidity.

As a subsequent step, we propose extending this analysis to different stock datasets to evaluate whether similar patterns of accuracy and feature importance persist, thereby validating the generalizability of our findings across varied market conditions.

\end{document}